\pdfoutput=1

\documentclass[%
 reprint,
superscriptaddress,
 amsmath,amssymb,
 aps,
prb,
showkeys,
floatfix,
]{revtex4-2}

\setcitestyle{super}

\usepackage{graphicx}
\usepackage{dcolumn}
\usepackage{bm}
\usepackage{siunitx}
\usepackage[version=4]{mhchem}
\usepackage{float}
\usepackage{physics}
\usepackage[hidelinks]{hyperref}  

\begin{document}

\preprint{APS/123-QED}

\title{Pressure dependence of intra- and interlayer excitons in 2H-\ce{MoS2} bilayers} 

\author{Paul Steeger}
\thanks{These two authors contributed equally}
\affiliation{University of Münster, Institute of Physics and Center for Nanotechnology, Wilhelm-Klemm-Str.\ 10, 48149 Münster, Germany}

\author{Jan-Hauke Graalmann}
\thanks{These two authors contributed equally}
\affiliation{University of Münster, Institute of Solid State Theory, Wilhelm-Klemm-Str.\ 10, 48149 Münster, Germany}

\author{Robert Schmidt}
\affiliation{University of Münster, Institute of Physics and Center for Nanotechnology, Wilhelm-Klemm-Str.\ 10, 48149 Münster, Germany}
\author{Ilya Kupenko}
\affiliation{University of Münster, Institute of Mineralogy, Corrensstr.\ 24, 48149 Münster, Germany}
\author{Carmen Sanchez-Valle}
\affiliation{University of Münster, Institute of Mineralogy, Corrensstr.\ 24, 48149 Münster, Germany}
\author{Philipp Marauhn}
\affiliation{University of Münster, Institute of Solid State Theory, Wilhelm-Klemm-Str.\ 10, 48149 Münster, Germany}
\author{Thorsten Deilmann}
\affiliation{University of Münster, Institute of Solid State Theory, Wilhelm-Klemm-Str.\ 10, 48149 Münster, Germany}
\author{Steffen Michaelis de Vasconcellos}
\affiliation{University of Münster, Institute of Physics and Center for Nanotechnology, Wilhelm-Klemm-Str.\ 10, 48149 Münster, Germany}
\author{Michael Rohlfing}
\affiliation{University of Münster, Institute of Solid State Theory, Wilhelm-Klemm-Str.\ 10, 48149 Münster, Germany}
\author{Rudolf Bratschitsch}
\email{Rudolf.Bratschitsch@uni-muenster.de}
\affiliation{University of Münster, Institute of Physics and Center for Nanotechnology, Wilhelm-Klemm-Str.\ 10, 48149 Münster, Germany}

\date{\today}

\begin{abstract}
The optical and electronic properties of multilayer transition metal dichalcogenides differ significantly from their monolayer counterparts due to interlayer interactions. The separation of individual layers can be tuned in a controlled way by applying pressure. Here, we use a diamond anvil cell to compress bilayers of 2H-\ce{MoS2} in the gigapascal range. By measuring optical transmission spectra, we find that increasing pressure leads to a decrease in the energy splitting between the A and interlayer exciton. Comparing our experimental findings with \textit{ab initio} calculations, we conclude that the observed changes are not due to the commonly assumed hydrostatic compression. This effect is attributed to the \ce{MoS2} bilayer adhering to the diamond, which reduces in-plane compression. Moreover, we demonstrate that the distinct real-space distributions and resulting contributions from the valence band account for the different pressure dependencies of the inter- and intralayer excitons in compressed \ce{MoS2} bilayers.



\end{abstract}

\keywords{MoS$_2$ bilayer, high pressure, diamond anvil cell, interlayer exciton, optical absorption}

\maketitle


\label{sec:maintext}
Molybdenum disulfide (\ce{MoS2}) is a layered van der Waals material belonging to the family of transition metal dichalcogenides (TMDCs).\cite{frisenda_naturally_2020} These materials exhibit diverse electronic band structures, ranging from metallic, semimetallic to semiconducting, depending on their chemical composition and crystal structure.\cite{kuc_electronic_2015, schwingenschloegl_heterostructures_2015} In its 2H-phase, bulk \ce{MoS2} is an indirect semiconductor. \cite{frindt_physical_1963} The discovery that \ce{MoS2} and other TMDCs act as direct semiconductors in their monolayer form has sparked significant interest in recent years.\cite{mak_atomically_2010, splendiani_emerging_2010, chiu_synthesis_2015, torche_biexcitons_2021}

Excitons, which are bound electron-hole pairs, play a crucial role in the opto-electronic response of \ce{MoS2} and similar materials, primarily due to their substantial binding energies. \cite{mueller_exciton_2018, he_tightly_2014} The surrounding environment strongly influences the behavior of excitons in monolayers, as demonstrated by depositing them on different substrates. The presence of a substrate enhances screening effects and induces spectral shifts in the exciton resonances.\cite{chernikov_exciton_2014, cadiz_excitonic_2017, ajayi_approaching_2017} Likewise, in heterostructures composed of two or more TMDC layers, the interaction between adjacent layers significantly affects the optoelectronic properties.\cite{novoselov_2D_2016, sun_excitons_2022, lukman_high_2020, latini_excitons_2015}. Notably, interlayer excitons can form, where the electron and hole reside in different layers. These interlayer excitons have been observed not only in heterostructures of TMDCs but also in homo-bilayers and bulk crystals, including \ce{MoS2}. \cite{rivera_interlayer_2018, deilmann_interlayer_2018, peimyoo_electrical_2021, cheng_role_2012, arora_interlayer_2017} 2H-\ce{MoS2} homo-bilayers are particularly  intriguing, because they exhibit interlayer excitons with a comparatively large oscillator strength and distinct energy separation from intralayer excitons.\cite{deilmann_interlayer_2018} As a consequence, the interlayer excitons are nicely discernible at room temperature, which renders 2H-\ce{MoS2} homo-bilayers an ideal model system for exploring interlayer interactions in a 2D semiconductor by optical spectroscopy.\cite{gerber_interlayer_2019, niehues_interlayer_2019} Using a diamond anvil cell (DAC), we are able to exert precise control over the interlayer distance in a systematic and continuous manner by applying high external pressures in the gigapascal range.\cite{pei_high_2022} We measure optical transmission spectra of bilayer 2H-\ce{MoS2} crystals from ambient pressure up to \SI{10}{GPa} and extract the energies of inter- and intralayer excitons. 

\begin{figure}[ht]
    \includegraphics{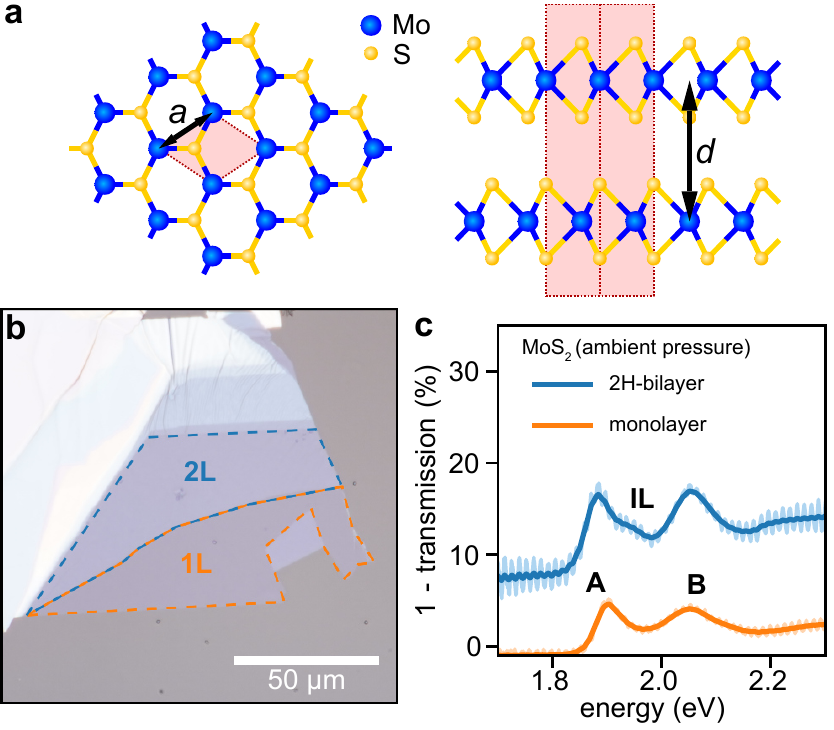}
    \caption{(a) Crystal structure of monolayer and 2H-bilayer \ce{MoS2} in top and side view. The unit cell is marked by the red shaded regions. $a$ and $d$ are the in-plane and out-of-plane lattice constants. (b) Optical micrograph of \ce{MoS2} crystals inside the diamond anvil cell. (c) Transmission spectra of monolayer and bilayer regions of \ce{MoS2} with the additional interlayer exciton resonance between A and B exciton in the bilayer. The data exhibits a high frequency oscillation caused by interference between the two diamond faces. A numerical low-pass filter yields the smooth lines shown here.}
    \label{fig:sample}
\end{figure}
\begin{figure*}[ht]
    \includegraphics{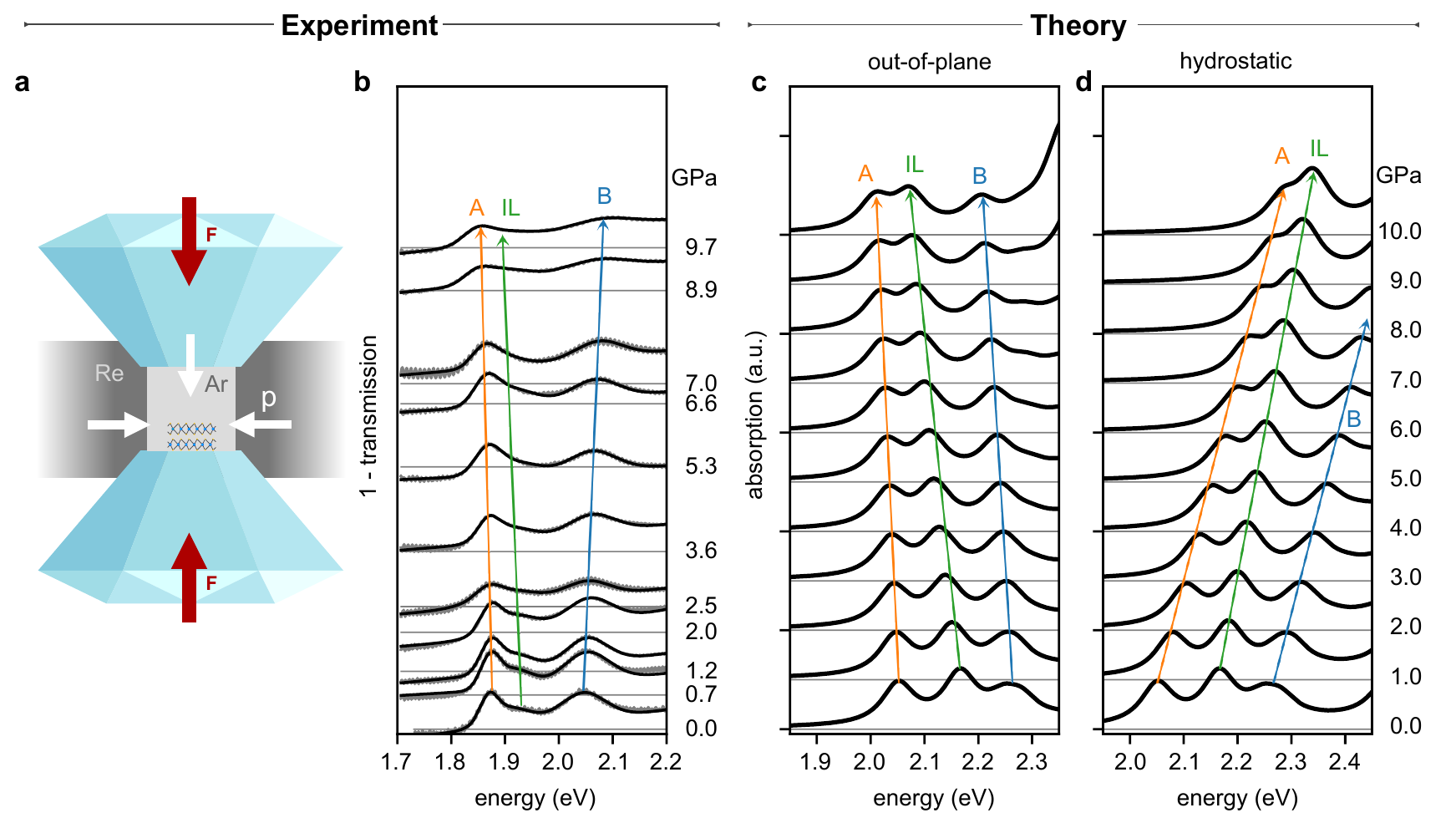}
    \caption{(a) Schematic drawing of the experiment. By pushing the two diamond anvils together, the argon inside the cell compresses the \ce{MoS2} bilayer, which resides on the bottom anvil. (b) Measured spectra (1-transmission) for different pressure values. The experimental data (gray) is fitted (see SI) with a model function (black) (c, d) Theoretical results for out-of-plane and hydrostatic compression. Exciton shifts are indicated with colored arrows. The two calculated extreme cases show a blue and red shift, respectively. The shifts in the experiment lie between these two cases, but mostly resemble the calculations for out-of-plane compression. The relative absorption strength and absolute energy positions of the different excitons are slightly different in experiment and theory.}
    \label{fig:spectra}
\end{figure*}

Figure~\ref{fig:sample}\,a depicts the crystal structure of mono- and bilayer \ce{MoS2}, illustrating the 2H stacking order of the crystals used in this study. In this polytype, consecutive layers are rotated by \ang{180} with respect to each other.\cite{wilson_transition_1969} After micromechanical exfoliation from a bulk crystal, potential \ce{MoS2} bilayers are identified by their contrast in the optical microscope and placed on the surface of one diamond anvil in a home-built microscopic stamping setup.\cite{castellanos-gomez_deterministic_2014} Figure~\ref{fig:sample}\,b displays an optical micrograph of \ce{MoS2} crystals with different layer numbers inside the diamond anvil cell, with distinctly different colors for monolayer and bilayer areas. The corresponding optical transmission spectra at ambient pressure and room temperature can be seen in Fig.~\ref{fig:sample}\,c. For the monolayer, the two main resonances, labelled A and B, are attributed to the two direct intralayer excitons at the K point, which are split in energy due to spin-orbit coupling (SOC).\cite{evans_optical_1965} The bilayer also exhibits A and B excitons with a slightly larger energy separation and an additional resonance in between, representing the interlayer (IL) exciton.\cite{gerber_interlayer_2019} The energy separations between these resonances are different for mono-, bi- and multilayer systems and can be used to confirm that the investigated sample is indeed a bilayer.\cite{niehues_interlayer_2019} The spectra in Fig.~\ref{fig:sample}\,c (and subsequent figures) are obtained by subtracting the measured wavelength-dependent transmittance of the sample (in \%) from 100\%, which resembles the absorption of the sample. Measured spectra of the \ce{MoS2} bilayer under different pressures are shown in Fig.~\ref{fig:spectra}\,b. As pressure increases, the three exciton resonances exhibit energy shifts, albeit with different rates. The A (B) exciton shifts linearly to lower (higher) energies with increasing pressure. This opposite trend results in an increased A-B splitting, indicating an increase in the valence band (VB) spin-orbit splitting (see SI for details). This effect is attributed to stronger interactions between layers, as they come closer under pressure, in agreement with previous studies.\cite{dou_probing_2016, hsu_quantitative_2022}

In Fig.~\ref{fig:spectra}, we present theoretically calculated absorption spectra (see SI for details) for pure out-of-plane compression (Fig.~\ref{fig:spectra}\,c) as well as hydrostatic pressure (Fig.~\ref{fig:spectra}\,d). When applying hydrostatic pressure to the bilayer, we observe a shift to higher energies, whereas out-of-plane compression results in a shift to lower energies. This contrasting behavior is caused by the response of the internal lattice parameters to pressure. 

In case of our out-of-plane calculations, we keep the lateral lattice constant $a$ fixed at its zero-pressure equilibrium value of \SI{3.18}{\AA}. The vertical compression reduces only the layer-to-layer distance $d$ and the thickness of the vacuum layer at a rate of about \SI{-0.10}{m\AA/GPa} at low pressure, gradually reducing to \SI{-0.03}{\AA/GPa} at higher pressure of \SI{10}{GPa}. Our geometric findings are consistent with previous works.\cite{fan_electronic_2015, peelaers_elastic_2014, yengejeh_effect_2020} As a consequence of this layer-to-layer compression, the direct KK gap of the band structure (see Fig. 3) shrinks from \SI{2.53}{eV} at zero pressure to \SI{2.43}{eV} at \SI{10}{GPa}, i.e.\ by (on average) \SI{-10}{meV/GPa}. Consequently, all optical transitions are redshifted upon increasing pressure.

\begin{figure}[ht]
    \includegraphics{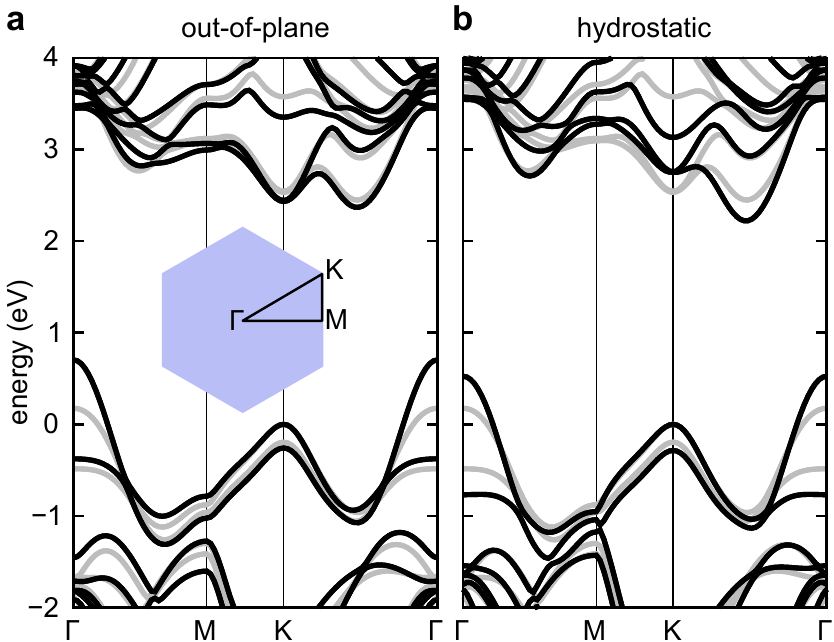}
    \caption{Calculated \textit{GdW} band structures of 2H-\ce{MoS2} at \SI{0}{GPa} (gray) and \SI{10}{GPa} (black) for out-of-plane (a) and hydrostatic (b) compression. The Brillouin zone with the path between the high symmetry points is shown in turquoise. All band structures are shifted so that the VB at the K point has an energy of \SI{0}{eV}.}
    \label{fig:bandstructure}
\end{figure}

In contrast, in the calculations for hydrostatic pressure we let the lateral lattice constant $a$ relax as well, at a rate of \SI{-7}{m\AA/GPa}. The effect of lateral (in-plane) deformation on the direct KK band gap is well known. While biaxial tensile strain leads to a red shift,\cite{plechinger_control_2015} in-plane compressive strain blue shifts all optical excitations, also for \ce{MoS2} monolayers.\cite{frisenda_biaxial_2017, heissenbuettel_nature_2019} In our case, the KK band gap increases by about \SI{30}{meV/GPa}, which overcompensates the gap reduction (\SI{-10}{meV/GPa}) of the simultaneous interlayer compression, such that in total the KK band gap grows from \SI{2.53}{eV} to \SI{2.75}{eV} under hydrostatic compression to \SI{10}{GPa}, i.e.\ by \SI{+22}{meV/GPa} on average (Fig.~\ref{fig:bandstructure}). This causes the strong blue shift of all optical excitations upon hydrostatic pressure (Fig.~\ref{fig:spectra}c). Note that the changes in $a$ are much weaker than those in $d$ by a factor of $\delta d/\delta a\approx3$, but the band structure is more sensitive to lateral in-plane than to out-of-plane compression, by a factor of 10. Lateral stress will thus dominate, if permitted.

Our calculated band structure data (Fig.~\ref{fig:bandstructure}) show that for both out-of-plane and hydrostatic compression, the VB splitting growths with pressure, while the CB splitting does not change significantly. As a consequence, the energy separation between the A and the B exciton increases upon pressure in both cases (out-of-plane and hydrostatic). Note that the shifts of the optical excitations as shown in Fig.~\ref{fig:spectra}\,c and d also include electron-hole interaction effects, and thus differ quantitatively from the bare band-structure arguments given above. 

Experimentally, we observe spectral shifts that lie between the two calculated extreme cases of entirely hydrostatic and pure out-of-plane compression. The TMDC crystal is attached to one of the diamond anvils of the pressure cell by van-der-Waals forces (Fig.~\ref{fig:spectra}\,a). If hydrostatic pressure is applied to the sample, not only the in-plane elastic modulus but also the adhesion between the crystal and the diamond anvil counteracts the in-plane component of the applied stress, resulting in an increased resistance of the sample to in-plane compression. As a result, the pressure-induced deformation of the crystal deviates from pure hydrostatic compression. The in-plane compression for a given applied pressure is therefore smaller than expected from elasticity theory. Quantifying this effect is challenging, as it varies from sample to sample, since the exact interaction between the sample and the anvil depends on sample size and interface purity. The absolute shift rates of individual resonances can lie somewhere between the two calculated cases (see also SI). 
However, comparing our theoretical results for the two extreme compression geometries with experiments on two different bilayers, we find that the relative shift rates, i.e., the pressure-dependent energy separations of the observed resonances, behave similarly in both compression scenarios and are hence reproducible. The interaction between the substrate and the 2D material samples inside the DAC has been previously observed.\cite{francisco-lopez_impact_2019, oliva_strong_2022,filintoglou_raman_2013} Our work provides an explanation for why the reported values for the pressure-dependent energy separation of the excitons in \ce{MoS2} bilayers are consistent not only in this study but also in earlier works, despite the deviations from hydrostatic compression.\cite{dou_probing_2016, hsu_quantitative_2022} While different interactions between sample and substrate lead to different shift rates of the individual excitons, the relative shift rates are very similar for any case between pure out-of-plane and completely hydrostatic compression. Furthermore, our findings are in agreement with measurements on bulk samples, \cite{brotons-gisbert_optical_2018, ci_quantifying_2017} which are less affected by the adhesion of the contact layer to the substrate and are therefore predominantly compressed hydrostatically.

\begin{figure}[htb]
    \includegraphics{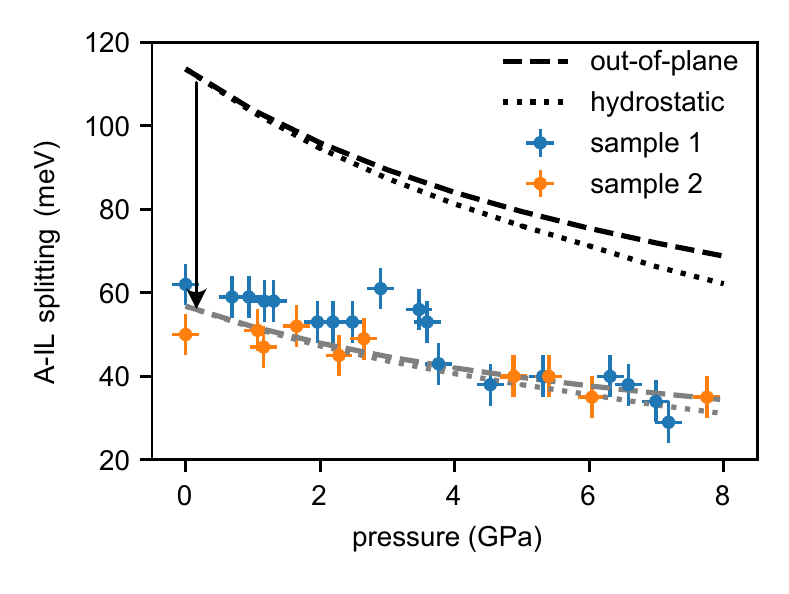}
    \caption{Energy splitting between interlayer (IL) and intralayer (A) excitons with increasing pressure. The experimental results of two samples in different DACs are compared to theoretical results (black) which are scaled down by a factor of 0.5 (gray) to compare the pressure dependence in experiment and theory.}
    \label{fig:EnergySplitting}
\end{figure}

We now turn our attention to the effect of crystal compression on the interlayer exciton. In earlier studies on 2H-\ce{MoS2} bilayers, the interlayer exciton was either not observed at all or hardly visible.\cite{dou_probing_2016, hsu_quantitative_2022, ci_quantifying_2017, li_tuning_2022, qiao_interlayer_2022} The measured transmission spectra (Fig.~\ref{fig:spectra}\,b) demonstrate that the IL exciton shifts to lower energies with increasing pressure and approaches the A exciton, indicating a difference in their shift rates. The pressure-dependent energy separation of the A and IL exciton is shown in Fig.~\ref{fig:EnergySplitting} for two \ce{MoS2} bilayer samples. At ambient pressure their separation is \SI{60}{meV} and it reduces to \SI{30}{meV} at \SI{8.9}{GPa}. 
The same trend is observed in both cases of the theoretical calculations, where we observe a reduction of the A-IL splitting, almost independent between hydrostatic and out-of-plane compression. However, we find that the calculated value for the A-IL splitting at ambient pressure is about twice as large as compared to the experiment. We attribute this mainly to small differences in the interlayer distance between theory and experiment due to the numerical implementation of the van der Waals interaction (D3 method by Grimme).\cite{grimme_consistent_2010} Furthermore, in theory, dielectric screening and binding to the substrate are not taken into account, because the calculations are performed for a sample in vacuum.\cite{greuling_spectral_2013} Therefore, we scale all values with a constant factor of two and observe that the pressure-induced reduction of the A-IL splitting matches well in theory and experiment.

\begin{figure}[ht]
    \includegraphics{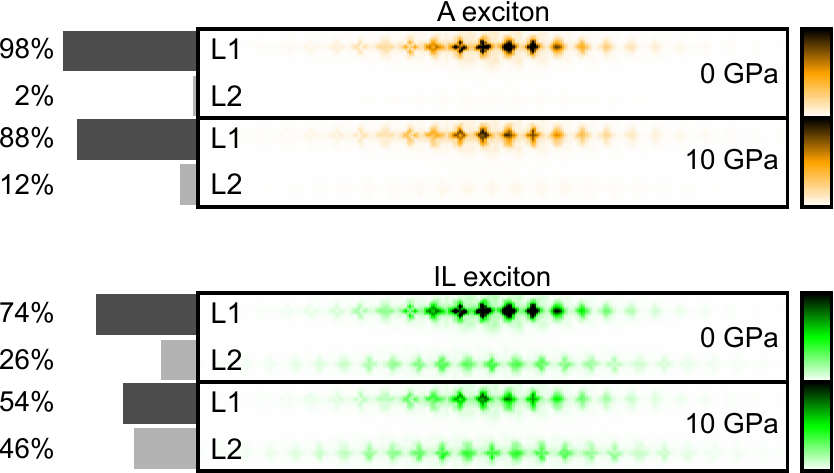}
    \caption{Real-space distributions of the A and IL exciton at ambient pressure and under out-of-plane compression. The color scale shows the probability density distribution of the excitons across two layers (L1, L2) in arbitrary units. The hole is fixed in the center of the upper layer, and the bar chart to the left of each distribution shows the integrated probability to find the electron in layer 1 (dark gray) or layer 2 (light gray). We observe an increasing interlayer contribution with increasing pressure, especially for the interlayer exciton.}
    \label{fig:RealSpaceContribution}
\end{figure}

Our calculations allow us to identify the reason why the interlayer (IL) exciton shifts differently under crystal compression than the intralayer (A) exciton. Both excitons originate from the K point of the Brillouin zone.\cite{gerber_interlayer_2019, niehues_interlayer_2019} Therefore, different shifts cannot be explained by changes of the gap between the uppermost VB and the lowermost CB. Also, the rate of the reduction of the binding energy is approximately the same for the A and IL exciton. We find that in addition to band gap and binding energy, pressure also influences the real-space distribution of A and IL excitons. They both exhibit an enhanced interlayer character (see Fig.~\ref{fig:RealSpaceContribution}). In case of the A exciton, the contribution of a transition from one layer to the other layer increases from a small \SI{2}{\%} to \SI{12}{\%} (\SI{7}{\%}) under \SI{10}{GPa} out-of-plane (hydrostatic) pressure. A similar effect can be observed for the IL exciton, which has an interlayer contribution at zero pressure of \SI{26}{\%}, and rises to \SI{40}{\%} (\SI{46}{\%}) at highest hydrostatic (out-of-plane) compression. The different spatial contributions also involve different bands. While the intralayer exciton is created by a transition from the uppermost VB to the lowermost CB, the interlayer exciton also shows contributions from the second-highest VB.\cite{deilmann_interlayer_2018} However, the latter shrinks due to the rising interlayer interaction from \SI{37}{\%} under \SI{0}{GPa} to \SI{9}{\%} under \SI{10}{GPa} out-of-plane pressure. We conclude that the decrease of the A-IL splitting is induced by the increase of the VB splitting (changes in the CB splitting only play a minor role), since the IL exciton is influenced stronger by this effect.

In conclusion, we have elucidated the effects of different types of compression on the optical absorption of bilayer 2H-\ce{MoS2} crystals by combining experimental measurements with \textit{ab initio} structural calculations. We find that due to the adhesion of the 2D material to the anvil, the in-plane compression is strongly reduced, resulting in a deviation from the usually assumed hydrostatic compression inside a diamond anvil cell. We observe a reduction of the energetic splitting between the intralayer A exciton and interlayer exciton under crystal compression, which we attribute to the contributions from the second-highest VB to the interlayer exciton. Our work highlights the importance of considering the complex interaction between the sample and substrate under pressure, and offers a framework for understanding experimentally observed pressure dependencies in the context of 2D materials inside a diamond anvil cell.

\begin{acknowledgements}
The project was funded by the Deutsche Forschungsgemeinschaft (DFG, German Research Foundation) through the priority program SPP2244 “2DMP” (Project-No 443405696) and through Project No. 426726249 (DE 2749/2-1 and DE 2749/2-2).
The authors gratefully acknowledge the Gauss Centre for Supercomputing e.V. (\url{www.gauss-centre.eu}) for funding this project by providing computing time through the John von Neumann Institute for Computing (NIC) on the GCS Supercomputer JUWELS \cite{alvarez_juwels_2021} at Jülich Super-computing Centre (JSC).
\end{acknowledgements}


\bibliography{main}

\clearpage
\widetext
\begin{center}
\textbf{\large Supporting Information}
\end{center}
\setcounter{equation}{0}
\setcounter{figure}{0}
\setcounter{table}{0}
\setcounter{page}{1}
\makeatletter
\renewcommand{\theequation}{S\arabic{equation}}
\renewcommand{\thefigure}{S\arabic{figure}}
\renewcommand{\bibnumfmt}[1]{[S#1]}
\renewcommand{\citenumfont}[1]{S#1}

\section{Experimental Methods}
Pressure-dependent optical experiments are conducted in a home-built piston-type diamond anvil cell (DAC) equipped with diamond anvils (type Ia, \SI{400}{\micro m} culet) mounted on tungsten carbide seats. The pressure is determined by measuring the energy shift of the R1 luminescence line of ruby microspheres inside the cell  \cite{shen_toward_2020}. The \ce{MoS2} crystals are mechanically exfoliated from bulk material (HQ Graphene) onto a polydimethylsiloxane (PDMS) polymer stamp. \ce{MoS2} flakes consisting of two or more layers are identified in the optical microscope by their optical contrast. They are then transferred from the PDMS stamp on top of one of the diamond anvils. Ruby spheres are placed next to the sample using a mechanical micro-manipulator. Afterwards, the DAC is immersed into liquid argon at ambient pressure and closed to seal the argon inside, which then acts as hydrostatic pressure transmitting medium during the experiment. 

Optical spectra are recorded at room temperature using a Czerny-Turner monochromator (Princeton Acton SP-2-500i) and a nitrogen-cooled CCD camera (Princeton PyLoN 100 BR eXcelon). Optical transmission spectra are obtained by focusing light from a white light-emitting diode (LED) onto the sample inside the DAC using an objective lens (Nikon TU Plan ELWD, 50x, NA 0.6), and collecting the transmitted light using a second objective lens (Mitutoyo M Plan Apo 50x, NA 0.55). The spectral signatures of the light source and the experimental setup are eliminated by corresponding reference measurements. These reference spectra are acquired through the DAC at a position adjacent to the sample. The measured spectra contain interference fringes resulting from reflections between the two opposing diamond faces. Since the diamonds are slightly tilted relative to each other, these fringes are wavelength-shifted for the sample and reference positions. Dividing the former by the latter introduces a non-vanishing artifact in the calculated transmittance values.

\section{Theoretical Methods}
The \textit{ab initio} calculations for the geometrical structure are based on the PBE functional within the density functional theory (DFT), including van der-Waals corrections (PBE+D3) \cite{grimme_consistent_2010}. We perform them with a basis set of localized Gaussian orbitals with three shells per atom. The corresponding decay constants (in a$_\text{B}^{-2}$) are [0.18, 0.49, 1.39] (\ce{Mo}) and [0.16, 0.56, 2.50] (\ce{S}). In order to integrate over the Brillouin zone, we use a $9\times9\times3$ k-point mesh. The structurally optimized in-plane lattice constant $a$ of the unstrained material amounts to \SI{3.18}{\AA}, the vertical \ce{Mo}-\ce{Mo} distance $d$ ($z$-direction) to \SI{6.18}{\AA}.

Linking the stress inside the DAC with the resulting strain of the sample is realized by using linear elasticity. To obtain the structure of the bilayer under pressure, we calculate the strain for each amount of pressure and apply it to the bulk crystal. The structurally optimized unit cell of the bulk crystal is then taken as the unit cell of the bilayer. For a hexagonal lattice under a certain stress $\vec{\sigma}=\pqty{\sigma_x,\sigma_y,\sigma_z}$, which is stressed additionally by $\Delta\vec{\sigma}=\pqty{\Delta\sigma_x,\Delta\sigma_y,\Delta\sigma_z}$, linear elasticity is given by
\begin{equation}
    \vec{\sigma}+\Delta\vec{\sigma}=\mqty(C_{11}\pqty{\vec{\sigma}}&C_{12}\pqty{\vec{\sigma}}&C_{13}\pqty{\vec{\sigma}}\\C_{12}\pqty{\vec{\sigma}}&C_{11}\pqty{\vec{\sigma}}&C_{13}\pqty{\vec{\sigma}}\\C_{13}\pqty{\vec{\sigma}}&C_{13}\pqty{\vec{\sigma}}&C_{33}\pqty{\vec{\sigma}})\cdot\vec{\varepsilon}.
    \label{hooke}
\end{equation}
Here, the strain $\vec{\varepsilon}=\pqty{\varepsilon_x,\varepsilon_y,\varepsilon_z}$ is related to the stress through the elasticity tensor, which consists of four independent elasticity parameters. These parameters are obtained by calculating the energy density $u$ via DFT for several small values of $\varepsilon_x$, $\varepsilon_y$ and $\varepsilon_z$ and fitting the data to
\begin{multline}
    u\pqty{\vec{\sigma},\vec{\varepsilon}}=u_0\pqty{\vec{\sigma}}+\frac{1}{2}\pqty{\varepsilon_x^2+\varepsilon_y^2}C_{11}\pqty{\vec{\sigma}}+\varepsilon_x\varepsilon_yC_{12}\pqty{\vec{\sigma}}\\+\pqty{\varepsilon_x+\varepsilon_y}\varepsilon_zC_{13}\pqty{\vec{\sigma}}+\frac{1}{2}\varepsilon_z^2C_{33}\pqty{\vec{\sigma}}-\frac{1}{2}\vec{\sigma}\vec{\varepsilon}.
    \label{energy}
\end{multline}
While applying hydrostatic pressure ($\Delta\sigma_x=\Delta\sigma_y=\Delta\sigma_z=-p$) leads to
\begin{equation}
    \dv{a\pqty{\vec{\sigma},p}}{p}=\frac{\pqty{C_{33}\pqty{\vec{\sigma}}-C_{13}\pqty{\vec{\sigma}}}a\pqty{\vec{\sigma},0}}{2C_{13}^2\pqty{\vec{\sigma}}-\pqty{C_{11}\pqty{\vec{\sigma}}+C_{12}\pqty{\vec{\sigma}}}C_{33}\pqty{\vec{\sigma}}}
    \label{hydroa}
\end{equation}
and
\begin{equation}
    \dv{d\pqty{\vec{\sigma},p}}{p}=\frac{\pqty{C_{11}\pqty{\vec{\sigma}}+C_{12}\pqty{\vec{\sigma}}-2C_{13}\pqty{\vec{\sigma}}}d\pqty{\vec{\sigma},0}}{2C_{13}^2\pqty{\vec{\sigma}}-\pqty{C_{11}\pqty{\vec{\sigma}}+C_{12}\pqty{\vec{\sigma}}}C_{33}\pqty{\vec{\sigma}}},
    \label{hydrod}
\end{equation}
out-of-plane pressure ($\varepsilon_x=\varepsilon_y=0$, $\Delta\sigma_z=-p$) results in
\begin{equation}
    \dv{d\pqty{\vec{\sigma},p}}{p}=-\frac{d\pqty{\vec{\sigma},0}}{C_{33}\pqty{\vec{\sigma}}}.
    \label{oopd}
\end{equation}
With knowing the values of $a$ and $d$ under zero stress and the derivatives of these quantities with respect to the applied pressure, we use the Runge-Kutta 4 method to calculate the structure up to \SI{10}{GPa} in steps of \SI{1}{GPa}. During each iteration, we fix the in-plane lattice constant as well as the interlayer distance and let the other atoms relax.

In order to get the absorption spectra, we use the \textit{GW}-BSE method within the LDA+\textit{GdW} approximation \cite{rohlfing_electronic_2010}. The dielectric screening is carried out by an atom-resolved model function, which is based on the random phase approximation. In order to represent the dielectric function and the screened Coulomb potential, we use a plane wave basis with a cut-off radius of \SI{2.5}{Ry}. The integration over the Brillouin zone is performed on a $k$-point mesh of $20\times20\times1$. Due to the screening of the periodically repeated bilayers, we extrapolate the band gap to an infinite vacuum size of the unit cell.

\newpage
\section{Fitting of measured optical spectra of two samples}
\begin{figure}[phtb]
    \centering
    \includegraphics{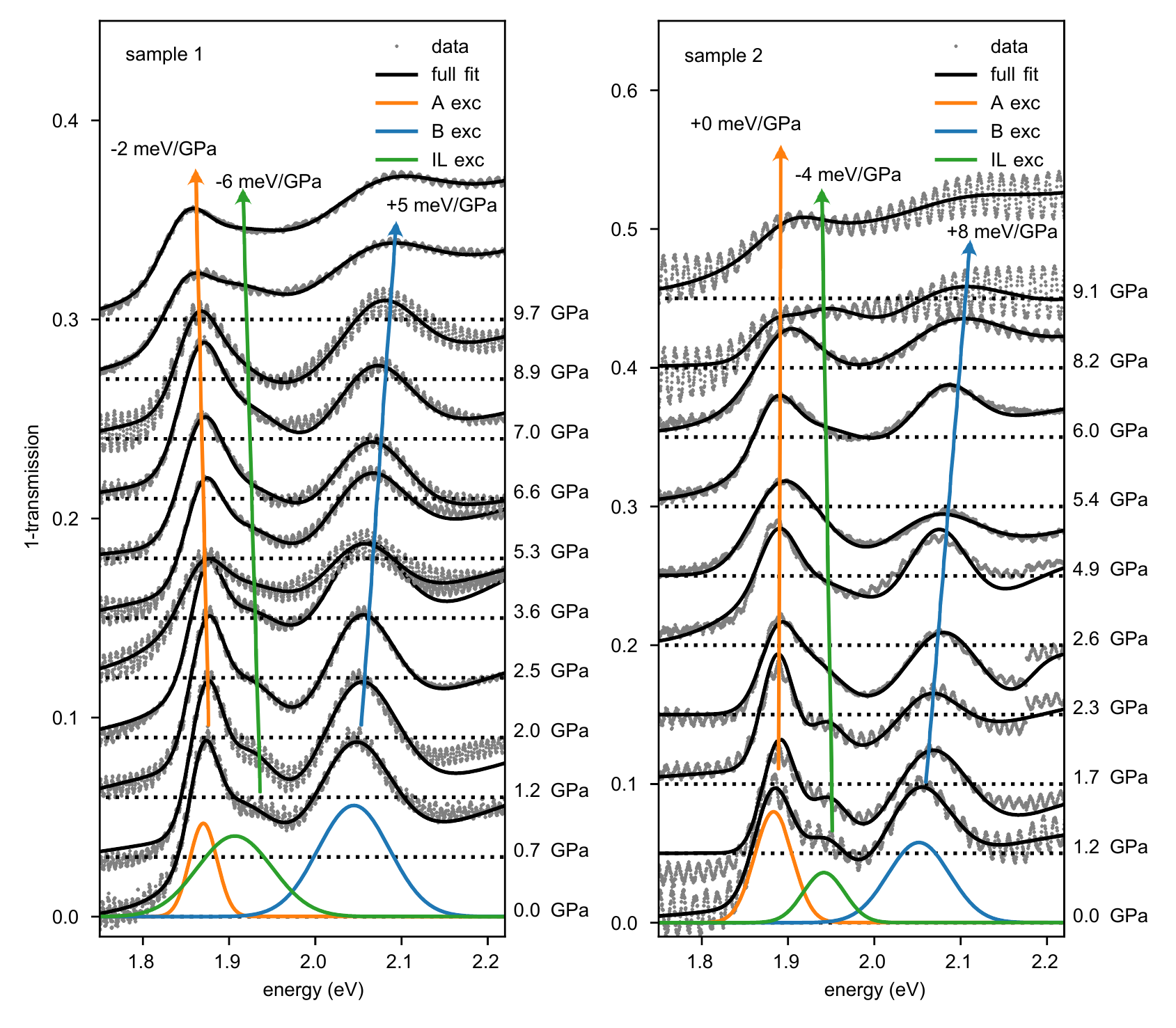}
    \caption{Measured pressure-dependent spectra for two different pressure series on different samples (sample 1, left; sample 2, right). The data is fitted using three Gaussian functions and a step-like function, rising towards higher energies to emulate broad higher-lying excitons. Spectra are offset vertically for clarity, and the individual excitonic resonances are shown exemplary for ambient pressure. Note that the interference in the data is not filtered numerically prior to the model fitting. From the extracted energy positions the shift rates of the individual excitons can be extracted. The values for sample 1 are closer to theoretical values for out-of-plane compression while sample 2 is closer to the hydrostatic case. Note, that relative shift rates are very similar.}
    \label{fig:fitting}
\end{figure}
\newpage

\section{Pressure-dependent energy separations}

\begin{figure}[phtb]
    \centering
    \includegraphics{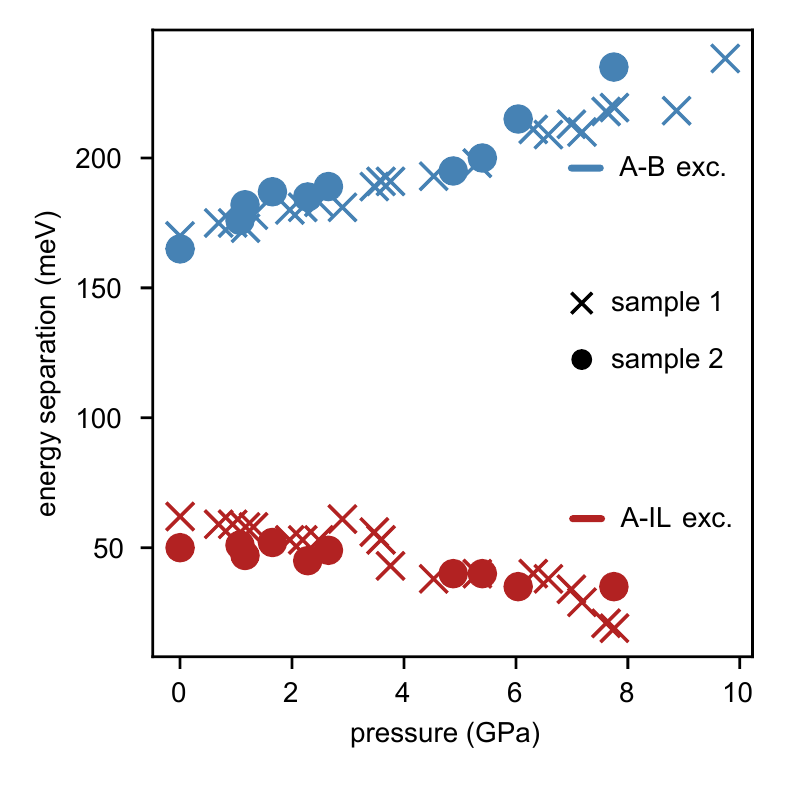}
    \caption{Energy separation between the observed exciton resonances under pressure for samples 1 and 2. The splitting of A and B excitons grows from \SI{170}{meV} at ambient pressure to \SI{240}{meV} at \SI{10}{GPa}, while the A-IL splitting decreases from \SI{60}{meV} to \SI{30}{meV} in the same pressure range.}
    \label{fig:splitting}
\end{figure}

\end{document}